\documentclass[referee]{raa}           
\usepackage{graphicx,times}
\usepackage{natbib}
\usepackage{amssymb,amsmath}
\bibpunct{(}{)}{;}{a}{}{,}

\usepackage[a4paper=true,dvipdfm=true,pagebackref=true]{hyperref}
\hypersetup{pdftitle = The title of my PDF, pdfauthor = My name, pdfsubject= The subject, pdfkeywords = keyword1 keyword2 keyword3} 
\hypersetup{colorlinks = true, linkcolor = green, anchorcolor = red, citecolor = blue, filecolor = red, pagecolor = red, urlcolor = red}

\begin{document}

   \title{Measuring the orbital periods of low mass X-ray binaries in the X-ray band
}

 \volnopage{ {\bf 2012} Vol.\ {\bf X} No. {\bf XX}, 000--000}
   \setcounter{page}{1}

   \author{Yi Chou\inst{}}

   \institute{ Graduate Institute of Astronomy, National Central University, Taiwan;  {\it yichou@astro.ncu.edu.tw}\\
\vs \no
}

\abstract{
A low mass X-ray binary (LMXB) contains either a neutron star or a  black hole accreting materials from its low mass companion star. It is one of the primary astrophysical sources for studying the stellar size compact objects and the accreting phenomena. As with other binary systems, the most important parameter of an LMXB is the orbital period, which allows us to learn about the nature of the binary system and constrain the properties of the system's components, including the compact object. As a result, measuring the orbital periods of LMXBs is essential for investigating these systems even though  fewer than half of them have known orbital periods. This article introduces the different methods for measuring the orbital periods in the X-ray band and reviews their application to various types of LMXBs, such as eclipsing and dipping sources, as well as pulsar LMXBs.  
}

   \authorrunning{Y. Chou }            
   \titlerunning{Measuring the orbital periods of the low mass X-ray binaries in X-ray band}  
   \maketitle

%
\section{Introduction}           
\label{sect:intro}

     Compact objects, white dwarfs, neutron stars and black holes, are the endings of the stellar evolution, and play important roles in the stellar life cycle. An X-ray binary contains either a neutron star or a black hole accreting material from a companion star and emits strong X-rays from the region close to the compact object. As a result, X-ray binaries are important astrophysical objects for studying stellar size black holes, neutron stars, as well as their accreting behavior. There are two types of X-ray binaries classified by their mass donors instead of accretors.  The companion of a low mass X-ray binary (LMXB) can be a late-type star or even a white dwarf or a brown dwarf with mass typically less than a solar mass. The two components of the binary are very close to each other so that the companion fills its Roche lobe and the material transfers through the inner Lagrangian point to the accretor side. Because the material carries a significant amount of angular momentum, an accretion disk is formed around the compact object. The material gradually loses its energy due to the viscosity and accretes onto the compact object. A large amount of  X-rays are emitted around the compact object because of the strong gravitational field.  In contrast, a high mass X-ray binary (HMXB) has a massive ($\geq$10 $M_{\sun}$) early-type star as the companion. The high mass companion loses its mass mainly through stellar wind and the accretor captures a large amount of matter in the close orbit, generating the X-rays.      
 
The variability of the X-ray binaries provides rich information for understanding the properties of these sources, including the nature of compact objects, accretion disks and the accretion process. The variability can be periodic, quasi-periodic or even aperiodic with time scales of less than a millisecond to thousands of days, difference of about 12 order of magnitudes. The variation amplitude may be as small as a few percent to a factor of thousands. Various time scale modulations allow us to study different parts of a binary system. For example, millisecond time scale variations are believed to be caused by the events in the region near the compact object, allowing us to study the compact object¡¦s properties. Conversely, some of long-term modulations may be caused by the accretion disk precession.

As with other binary systems, the orbital period, which directly reflects the binary nature, is the most fundamental parameter for an X-ray binary. The orbital periods of the X-ray binaries range from less than an hour to hundreds of days. The HMXBs usually have longer orbital periods, from days to hundreds of days. The LMXBs are more compact, and their orbital periods range from tens of minutes to several hours. The binary size, i.e. the separation of the two components ($a$), can be estimated using Kepler's third law,
 
\begin{eqnarray}\label{k3l}
{a \over R_{\sun} }& = &9.03 \Bigl({{m_a+m_d} \over {10 M_{\sun}}} \Bigr)^{1/3}{\Bigl({P_{orb} \over {1\,d}}\Bigr)}^{2/3}  \\
\nonumber & = & 0.50 \Bigl({{m_a+m_d} \over {M_{\sun}}} \Bigr)^{1/3}{\Bigl({P_{orb} \over {1\,h}}\Bigr)}^{2/3}
\end{eqnarray}

\noindent where $m_a$ and $m_d$ are the masses of the accretor and donor respectively. Equation~\ref{k3l} shows the compactness of the X-ray binary systems; the binary size of a HMXB is about tens of solar radii but the binary size of an LMXB is typically less than or equal to a solar radius.

However, fewer than half of the X-ray binaries have known orbital periods. According to the 4th edition of catalogs of HMXB (\citealt{Liu+etal+2006}) and LMXB (\citealt{Liu+etal+2007}),  only 45 out of 114 HMXBs and 74 out of 187 LMXBs have recorded orbital periods. These orbital periods are obtained from either optical or X-ray bands. Because the X-ray binaries are rather compact, it is almost impossible to resolve their orbits in images. Therefore, the measurements of the orbital periods are primarily achieved by variations in light curves or radial velocities. In most cases, the orbital variations are easier to detect for systems with high inclination angles. Probably the only exception is the Be X-ray binary whose orbital period can be determined by recurring X-ray outbursts due to the neutron star periodically passing through the disk around the Be star and accreting large amounts of gas from it.     

The orbital periods of many X-ray binaries are measured by their optical emissions. These can be obtained either from the periodic flux modulations or the orbital Doppler effect of the companion stars. The optical emissions for a HMXB are mostly from its mass donor with little affected by the X-ray source. Conversely, for an LMXB, both the companion star and accretion disk may contribute to the optical emissions. For persistent LMXBs or transient LMXBs during the outburst state, the optical emissions are dominated by the accretion disks, due to reprocessing X-rays. The optical emissions from the accretion disk can be asymmetric or periodically blocked by the companion. However, the optical modulation from an accretion disk may be coupled with the disk motion so that the variation period can be different from the orbital period, which will be further discussed in Section ~\ref{sect:DXS}. For a transient LMXB during the quiescent state, the optical emissions are from the companion. Because the companion is tidally distorted, the emissions are also asymmetric because of  gravitational darkening and the ellipsoidal modulation is observed.  Another way to detect the orbital period in the optical band, like the single-lined spectroscopic binary, is to measure the radial velocity variation of the companion star.  The radial velocity variation provides not only the orbital period, but also other orbital parameters, such as the projected semi-major axis ($a \sin i$) of the companion's orbit and eccentricity ($e$). Furthermore, the mass function

\begin{equation}
f(m)= {{m_a \sin^3 i} \over {(m_a + m_d)}} = {{P_{orb}K_d^3(1-e^2)^{3/2}} \over {2 \pi G}}
\end{equation}
\noindent where $K_d$ is the semi-amplitude of the radial velocity of the companion, allows us to constrain the mass of the compact object. 

However, there are still some difficulties with measuring the orbital periods in the \ optical band. First, some X-ray binaries have no identified optical counterparts. The galactic HMXBs are mostly distributed around the galactic plane where the extinction for the optical band is high.  For the LMXBs, some are located in the globular clusters so their optical counterparts are difficult to identified in such star crowded region. Even if the optical counterparts are known, measuring the orbital Doppler effects from the companion for some LMXBs are still very difficult because they are too dim.  An alternative way to probe the orbital periods of X-ray binaries is from their X-ray emissions. This article focuses on measuring the orbital periods of LMXBs. There are two major methods of measuring the orbital periods of LMXBs in the X-ray band: from their X-ray flux variations or from pulsations of neutron stars. Measuring the orbital period with the X-ray flux is introduced in Section ~\ref{sect:OMX}, including general period search methods (Section ~\ref{sect:PSM}), the O-C method (Section ~\ref{sect:OCM}), and further discussion on the eclipsing sources (Section ~\ref{sect:EXS}) and dipping sources (Section ~\ref{sect:DXS}). It also covers other variations related to the orbital period measurements, such as the accretion disk dynamics. Measurements of LMXB orbital periods using the pulsations from neutron stars is described in Section ~\ref{sect:OXP}.



\section{Measuring the orbital period with X-ray flux modulation}
\label{sect:OMX}

\begin{figure}
   \centering
   \includegraphics[width=14.0cm, angle=0]{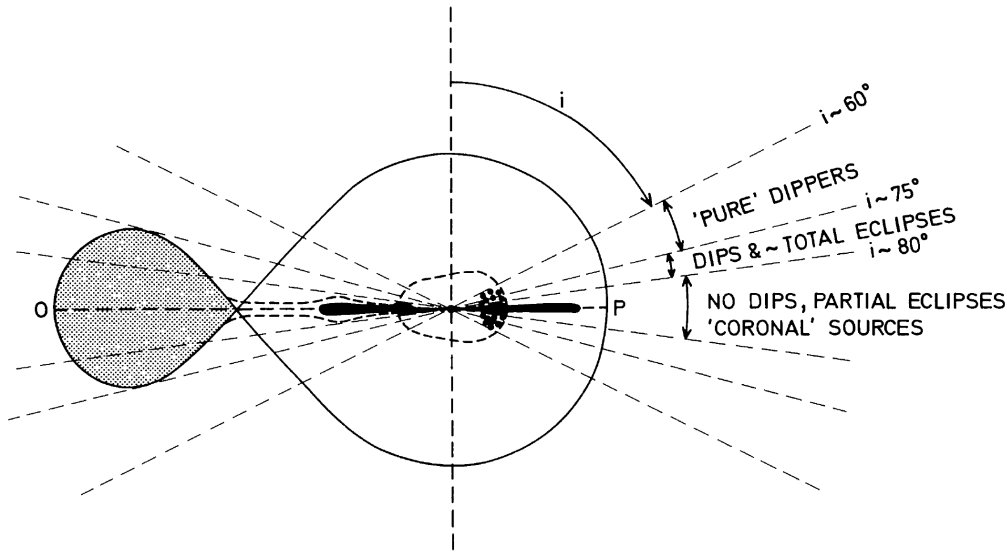}
   \caption{Various types of X-ray flux modulations from orbital motions of LMXBs caused by the different inclination angles. Adopted from \cite{Frank+etal+1987}} 
   \label{frank1987}
   \end{figure}

One of the primary methods of measuring the orbital period of an LMXB is from its X-ray flux variation. The X-rays emitted from the region near the compact object are periodically obscured, either fully or partially by the companion, or absorbed by the outer structure of the accretion disk, which causes the orbital modulations in X-ray band. Such modulations can be seen only for a binary system with high orbital inclination angle ($i$). According to \cite{Frank+etal+1987} (see Figure~\ref{frank1987}), no orbital variation can be observed for an LMXB system with orbital inclination angle less than $60{\degr}$. For a system with orbital inclination angle between $60{\degr}$ and $75{\degr}$, periodic X-ray dips caused by the absorption of X-rays by the vertical structure on the accretion disk can be detected in the X-ray light curve. If the inclination angle is larger than $75{\degr}$, an eclipse occurs. Interestingly, the total eclipse, together with the dip, is seen for orbital inclination between $75{\degr}$ and $80{\degr}$ if the X-ray emission region is point-like. In principle, no X-rays can be detected for higher inclination angle because the X-ray source is completely blocked by the finite thickness of the accretion disk. However, if the X-ray emission region is extended due to the X-rays being scattered by the accretion disk corona (ADC), the companion can only occult part of the X-ray emission region and result in a partial eclipse (\citealt{White+Holt+1982}). A typical partial eclipsing X-ray binary is X1822-371 (\citealt{White+etal+1981}).  

\subsection{Period Search Methods}
\label{sect:PSM}

Because the orbital modulation is periodic, the orbital period can be obtained by various period search methods, such as Lomb-Scargle periodogram (hereafter LS periodogram, \citealt{Lomb+1976,Scargle+1982}) and phase dispersion minimization (PDM, \citealt{Stellingwerf+1978}). The best method for searching the orbital period depends on the modulation profile. For example, the LS periodogram, based on fitting a sinusoidal function to the light curve, is more sensitive to smooth variations.  For a highly non-sinusoidal profile, such as the X-ray dip (Section~\ref{sect:DXS}) with a short duty cycle, PDM is a better choice because the LS power of the signal would be spread out to its harmonics and therefore be buried by the noise in the power spectrum.  Different period search methods are usually applied to confirm the periodicity, especially if the modulation profile is unknown prior to measurement. Many period search methods, such as the LS periodogram, must be applied to the binned light curve. Furthermore, for the methods where the period is obtained by an index made by the folded light curve folded with a trial period, such as PDM, the bin size in the folded light curve must be specified. Consequently, results may be slightly different with different bin sizes for the same light curve. However, the X-ray data can be recorded as an event list that provides the event arrival times for the detected X-ray photons. The Z-test (\citealt{Buccheri+etal+1983}) and H-test (\citealt{deJager+etal+1989}) directly operate on the event arrival time without binning and a unique result can be obtained.

In addition to finding the orbital period, the primary purpose for a period search is to verify the significance of the signal.  A periodic signal results high power and exhibits a peak in the power spectrum (or a local minimum in the PDM spectrum). However, the peak may be caused by noise although this probability is small. The significance can be evaluated by hypothesis testing. The null hypothesis is that there is no periodic signal in the light curve, so that all variations are due to noise. The null hypothesis would be rejected if the probability that the peak is due to noise (also called the false alarm probability) is smaller than a prior significance level. Taking the LS periodogram as an example, the probability distribution of pure white noise in the power spectrum is an exponential distribution with a mean value of 1. A peak with power greater than 6 indicates that the probability that the peak is caused by noise is less than $exp(-6.0)=0.0025$. In other words, we have a more than 99.75\% (3$\sigma$) confidence level for the detection of periodic signal.

Once the orbital period is determined by the period search method, the error of the period must also be specified. The uncertainty of the frequency is sometimes estimated as the width of the peak in the power spectrum but this value can only be considered as the upper limit of the uncertainty (\citealt{Levine+etal+2011}). The width of the peak is inversely proportional to the time span of the light curve and independent of the power of the peak. However, we expect that in principle, the uncertainty of the period is smaller for a more significant signal. Taking LS periodogram as example, from \cite{Horne+Baliunas+1986}, the uncertainty in frequency may be estimated as          
\begin{equation}
\delta f={3 \over {8T \sqrt{P}}}
\end{equation}

\noindent where $T$ is the time span of the light curve and $P$ is the power of the signal. For a significant detection of a periodic signal, the LS power is much larger than 1 so the frequency uncertainty is much smaller than $1/T$. For a period search method where period uncertainty is hard to be determined, Monte Carlo simulation can be used to estimate the uncertainty.

Like optical observations, X-ray light curves are usually unevenly sampled so the discrete Fourier transfer is improper for searching the periodicity. In contrast, the LS periodogram, which is based on fitting a sinusoidal function, is suitable for unevenly sampled data. However, all period search methods suffer from  large gaps in the light curves. For X-ray observations, the gaps may be caused by the Earth occultation, which is more severe for low orbit satellites, or the observation being terminated because the target is too close to the Sun. These observation gaps usually result in aliases around the true signal. If the gaps are periodic, like the Earth occultation, whose period is the satellite orbital period, the aliases appear to be equally spaced around the signal on the spectrum where the frequency difference equals to the frequency of the gaps on the light curve. Furthermore, if the gaps are too large, the aliases may be more powerful than the true signal in the power spectrum, giving incorrect period of the light curve.  

\subsection{O-C Methods}
\label{sect:OCM}
An alternative way to obtain or refine the orbital period is the O-C (observed minus calculated) method. The O-C method usually can provide a more precise orbital period measurement, about an order of magnitude better than the period search methods given in Section~\ref{sect:PSM}.  Furthermore, it is able to probe the orbital period derivative (i.e. $\dot P_{orb}$, $\ddot P_{orb}$ etc.) To apply this method, a relatively stable orbital modulation profile is preferred, allowing a specific reference point, called the fiducial point, in the profile to be well-defined. The O-C method traces the change in the occurrence times of the fiducial points.  For a given reference time $T_0$ and expanding the orbital frequency ($\nu_{orb} (t)$) as a polynomial, the observed cycle count ($N_{obs}$) can be expressed as
  
 \begin{eqnarray}\label{oc1}
N_{obs}(t) &=& \int_{T_0}^t \nu_{orb} (t^{\prime}) dt^{\prime}=\int_{T_0}^t (\nu_{orb,0}+\dot \nu_{orb,0}(t-T_0)+ {1 \over 2} \ddot \nu_{orb,0}(t-T_0)^2 +...... )dt^{\prime}\\
\nonumber & = & N_{obs}(T_0)+\nu_{orb,0}(t-T_0)+{1 \over 2} \dot \nu_{orb,0}(t-T_0)^2+{1 \over 6} \ddot \nu_{orb,0}(t-T_0)^3 +......
\end{eqnarray}

\noindent The reference time, usually called the phase zero epoch, is usually chosen as an occurrence time of a fiducial point so that $N_{obs}(T_0)=0$. However, the observed cycle count number increases rapidly with time so that it is almost always compared with a ``calculated''  cycle count number. The linear ephemeris for the calculated cycle count is usually chosen as

\begin{equation}\label{oc2}
N_c (t)=N_c (T_0^{\prime})+\nu_{c,0}(t-T_0^{\prime})
\end{equation}

\noindent The frequency ($\nu_{c,0}$), sometimes called folding frequency, may be obtained by the period search methods discussed above or using a previously reported value. The difference in the cycle count can be written as
  
\begin{eqnarray}\label{oc3}
\phi(t) &=& N_c(t)-N_{obs} (t)\\
\nonumber        &=& N_c (T_0^{\prime})+\nu_{orb,0}(t-T_0^{\prime})\\
\nonumber        & &-[N_{obs}(T_0)+\nu_{orb,0}(t-T_0)+{1 \over 2} \dot \nu_{orb,0}(t-T_0)^2+{1 \over 6} \ddot \nu_{orb,0}(t-T_0)^3 ]\\
\nonumber &=&-[\nu_{orb,0}(T_0^{\prime}-T_0)+{1 \over 2}\dot \nu_{orb,0}(T_0^{\prime}-T_0)^2+{1 \over 6} \ddot \nu_{orb,0}(T_0^{\prime}-T_0)^3]\\
\nonumber & &+[\nu_{c,0}-\nu_{orb,0}-\dot \nu_{orb,0}(T_0^{\prime}-T_0)-{1 \over 2}\ddot \nu_{orb,0}(T_0^{\prime}-T_0)^2](t-T_0^{\prime}) \\
\nonumber & &-{1 \over 2} [\dot \nu_{orb,0}+\ddot \nu_{orb,0}(T_0^{\prime}-T_0)](t-T_0^{\prime})^2-{1 \over 6} \ddot \nu_{orb,0}(t-T_0^{\prime})^3
\end{eqnarray}

\noindent where $N_c (T_0^{\prime})=N_{obs}(T_0)=0$ if $T_0^{\prime}$ and $T_0$ are the phase zero epochs for calculated and observed cycle counts respectively. $\phi(t)$ is called the phase and usually less than one cycle. The phase can be obtained by the location of the fiducial point on the folded light curve. If both $T_0^{\prime}$ and $\nu_{c,0}$ are close to their corresponding true values $T_0$ and $\nu_{orb,0}$ and the higher-order terms are small, the phase drift can be continuously traced. The orbital parameters are obtained by fitting a polynomial to the evolution of the phases of fiducial points from the data. Using the period as a parameter, the Equation~\ref{oc3} can be approximately written as (up to quadratic terms) as
   \begin{eqnarray}\label{oc4}
\phi(t) &\approx&\phi_0 +{{P_{orb,0}-P_{c,0}} \over {P_{orb,0}P_{c,0}}}(t-T_0^{\prime}) +{1 \over 2}  {{\dot P_{orb}} \over {P_{orb,0}^2}}(t-T_0^{\prime})^2
\end{eqnarray}

The phase of the fiducial point is sometimes, obtained by the folded light curve folded by a linear ephemeris (i.e. $T_N^{\prime}=T_0^{\prime}+P_{c,0}N$), particular for an orbital profile that is unclear for a single orbital cycle in a light curve. However, cycle count ambiguity must be treat with caution. The measured phase shift $\phi$ could actually be $(\phi+/-N^{\prime})$ where $N^{\prime}$ is an integer. This could happen if the folding period (or frequency)  deviates by a large amount, perhaps due to large error, from the true orbital period and also the large gap in data. Phase uncertainty due to the folding period deviation or error would accumulate with time and result in cycle count ambiguity. Therefore, a small uncertainty in the folding period is essential from extrapolating the folding ephemeris to larger time spans, especially if there are large gaps in the data. More detailed discussions about this issue can be found in \citet{Chou+etal+2001}.

The orbital ephemeris is usually written as a function of cycle count. Taking the quadratic ephemeris as an example, the time of the fiducial point at the Nth cycle is expressed as
      
\begin{equation}
T_N \approx T_0+P_{orb,0}N+{1 \over 2}P_{orb,0} \dot P_{orb} N^2
\end{equation}
\noindent Thus, using the O-C method, the time delay in comparison to the ``calculated'' linear ephemeris, $T_N^{\prime} =T_0^{\prime} + P_{c,0}N$, is

\begin{equation}
\Delta t=T_N - T_N^{\prime} \approx (T_0 - T_0^{\prime})+(P_{orb,0}-P_{c,0})N+{1 \over 2}P_{orb,0} \dot P_{orb} N^2
\end{equation}

\noindent The parameters can be obtained or refined by fitting a polynomial to the time delay as a function of cycle count. The application of the O-C method to measure the orbital period and its derivative will be further discussed in the following sections for eclipsing and dipping LMXBs.

\subsection{Eclipsing X-ray Sources}
\label{sect:EXS}
\begin{figure}
   \centering
   \includegraphics[width=14.0cm, angle=0]{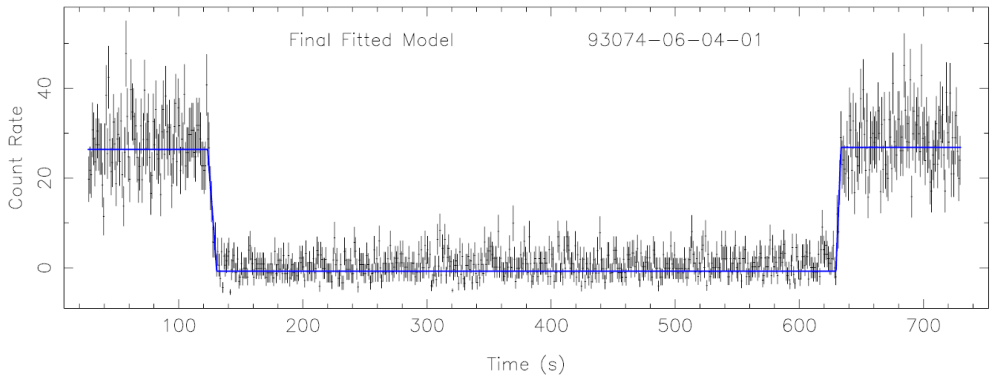}
   \caption{A typical total eclipsing profile in the X-ray band detected from EXO 0748-676.  Adopted from \cite{Wolff+etal+2009}} 
   \label{wollf2009-0}
   \end{figure}

In addition to the radial velocity variation from the orbital Doppler effect, the most direct evidence for binary orbital motion is eclipsing. For an X-ray binary with sufficiently high inclination angle, the X-ray emissions may be totally blocked by the companion if the X-ray emission region is point-like and exhibits a total eclipsing profile. A typical X-ray total eclipsing profile is shown in Figure~\ref{wollf2009-0} Because the total eclipsing profile can be very well defined, the O-C method is usually applied to refine the orbital period or trace the orbital period change. The most commonly used fiducial point is the mid of eclipse time obtained by fitting a defined model to the profile.  The simple step and ramp model is usually adopted to fit the eclipsing profile (\citealt{Parmar+etal+1991,Hertz+etal+1997}).   

\begin{figure}
   \centering
   \includegraphics[width=14.0cm, angle=0]{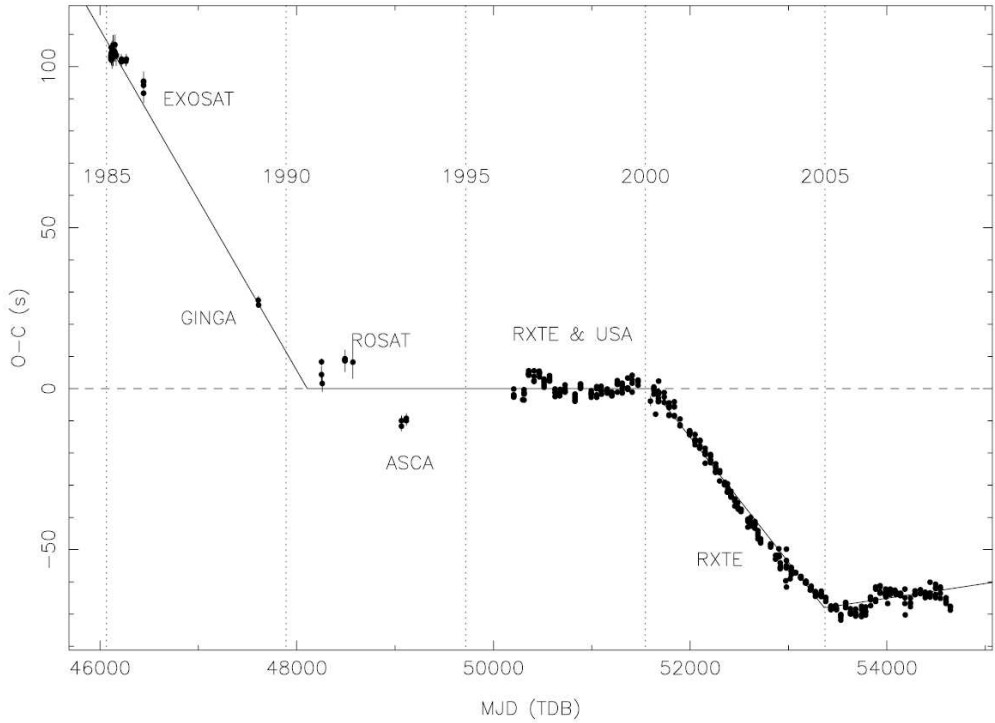}
   \caption{The O-C results from over 20 years' of monitoring the mid of eclipse times of  EXO 0748-676. The orbital period glitches can be clearly seen. Adopted from  \cite{Wolff+etal+2009}} 
   \label{wollf2009-1}
   \end{figure}

To date, total X-ray eclipses have been detected in 12 LMXBs (EXO 0748-676, Her X-1, X 1658-298, XTE J1710-281, AX J1745.6-2901, GRS 1747-312, CXOGC J174540.0-290031, Swift J1749.4-2807, 1RXS J175721.2-304405, 4U 2129+47 and two quiescient sources, X5 and W37, in globular cluster 47 Tuc (\citealt{Heinke+etal+2003,Heinke+etal+2005})). For the newly discovered total eclipsing source 1RXS J175721.2-304405, only a part of the eclipsing profile was observed so its orbital period remains unknown (\citealt{Maeda+etal+2013}). The 4U 2129+47 used to be recognized as a typical partial eclipsing source but a total eclipsing profile has been detected during its quiescent state (\citealt{Nowak+etal+2002}). Among these 7 total eclipsing LMXBs, the most well-studied is EXO 0748-676 , a transient source with a neutron star as its accretor. The eclipse timing has been traced since its discovery in February 1985. However, the phases of the mid of eclipse times behave irregularly (see Figure~\ref{wollf2009-1}) and no simple ephemeris can fit the O-C residuals. It appears that the orbital period has abruptly changed several times by an order of milliseconds over more than 20 years of monitoring (\citealt{Wolff+etal+2009}). A similar phenomenon was also seen in XTE J1710-281 (\citealt{Jain+Paul+2011}). This orbital period glitch is likely due to the strong magnetic activity which changes the gravitational quadrupole moment of the companion star (\citealt{Wolff+etal+2007,Wolff+etal+2009,Jain+Paul+2011})

\begin{figure}
   \centering
   \includegraphics[width=14.0cm, angle=0]{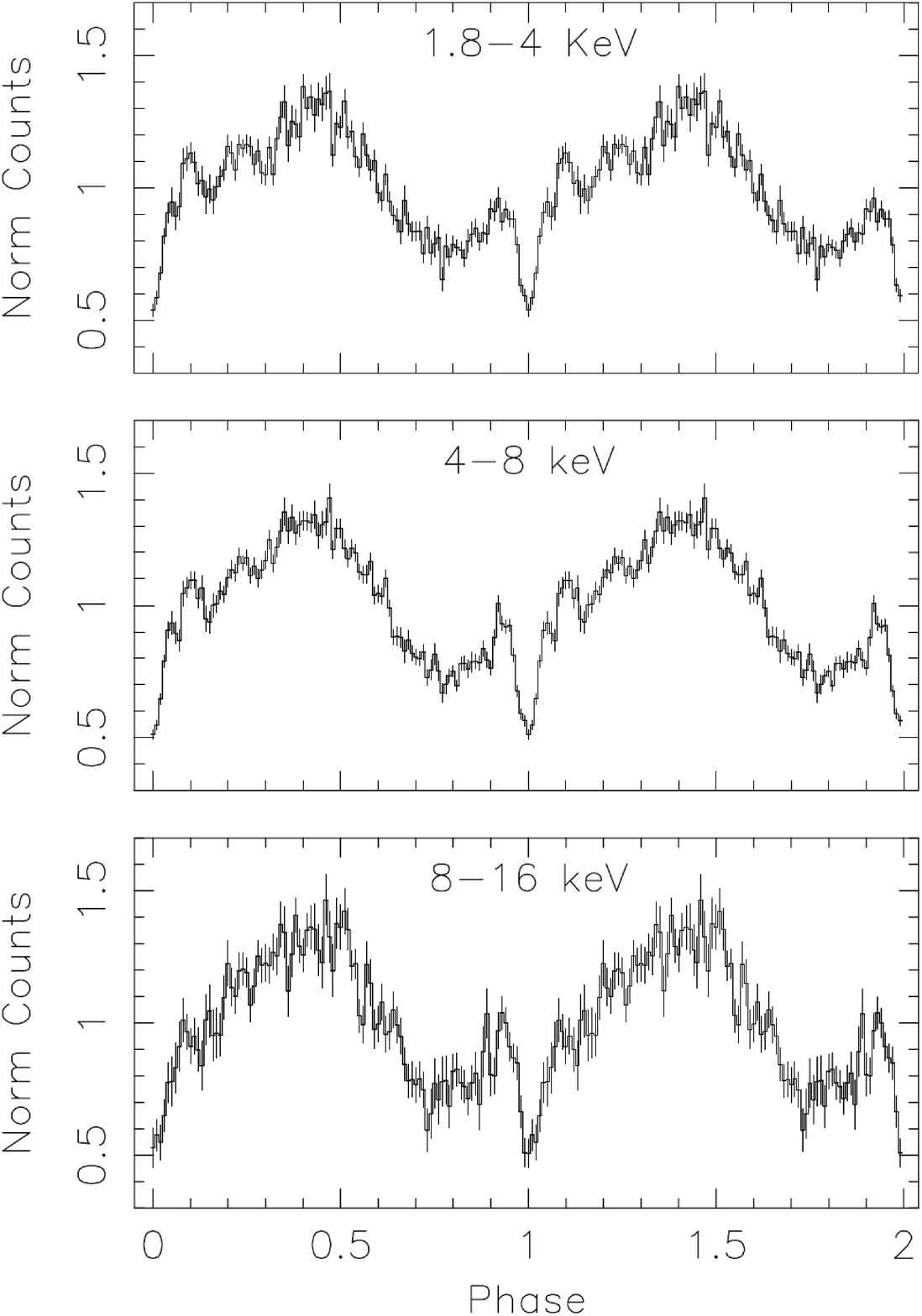}
   \caption{A typical partial eclipsing profile in the X-ray band detected from   4U 1822-37. Adopted from \cite{Parmar+etal+2000}} 
   \label{parmar2000}
   \end{figure}

There are 5 LMXBs exhibiting partial X-ray eclipses (2S 0921-630, 4U 1822-37, XTE 2123-056, 4U 2129+12 and 4U 2129+47).  The partial X-ray eclipse provides evidence that the X-ray emission region is extended due to the X-rays being scattered in the ADC (\citealt{White+Holt+1982}).  For an ADC source, the value of $L_x/L_{opt}$ is about 20 whereas this value is about 100-1000 for the other LMXBs. 4U 1822-37 is one of the best-studied partial eclipsing LMXB systems.  However, the orbital modulation profile is not purely a partial eclipse. In addition to the partial eclipse profile, there is a smooth variation that reaches its minimum at about 0.2 cycles before the eclipse (\citealt{Parmar+etal+2000}, see Figure~\ref{parmar2000}). This smooth variation is caused by the X-rays from ADC being obscured by the structure at the rim of the accretion disk (\citealt{White+Holt+1982}). The minimum of the partial eclipse is usually chosen as the fiducial point (also called the arrival time of the eclipse). Its value is obtained by fitting a Gaussian to the partial eclipse profile, plus a function (e.g. a polynomial, sinusoidal function) to model the smooth modulation (\citealt{Iaria+etal+2011}). After more than 30 years of tracing the orbital evolution, the orbital period change rate was reported to be $1.59(9) \times 10^{-10}$ s s$^{-1}$, which is three orders of magnitude greater than the theoretical prediction based on the mass conservation in the binary system (\citealt{Iaria+etal+2011}). On the other hand, 4U 1822-37 is also an X-ray pulsar with the neutron star spin period of 0.59 s (\citealt{Jonker+vanderKlis+2001}), which provides an alternative way to independently probe the orbital period evolution. 

\subsection{Dipping X-ray Sources}
\label{sect:DXS}
\begin{figure}
   \centering
   \includegraphics[width=14.0cm, angle=0]{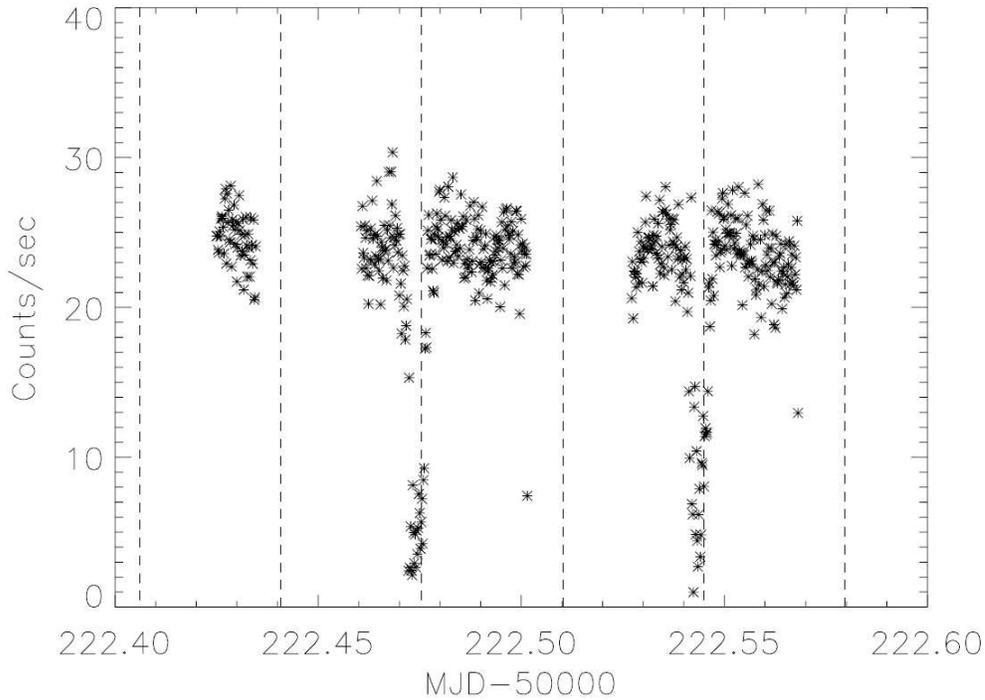}
   \caption{A typical X-ray dip profile detected from X 1916-053. The dashed lines are the expected dip center times. Adopted from \cite{Chou+etal+2001}} 
   \label{chou2001}
   \end{figure}

In addition to eclipsing sources, some LMXBs exhibit orbital modulations in X-ray band that are believed to be due to absorption of X-rays by the accretion disk structure. The typical one is the dipping source. X-ray dips have been seen in more than a dozen LMXBs, where the X-ray intensity periodically decreases by a factor with duration from $<$0.1 up to ~0.4 duty cycles. Figure~\ref{chou2001} shows a typical light curve of a dipping LMXB. This is caused by the absorption of X-rays emitted from the region around the compact object by the vertical structure in the outer part of the accretion disk. Therefore, the period of the dip is considered to be the orbital period of the binary system. X-ray dips have been seen in total eclipsing LMXBs (e.g. EXO 0748-676) and partial eclipsing sources (e.g. 4U 2129+12). However, there are a number of LMXBs exhibiting only pure dips without an eclipse, classified as dipping LMXBs. In contrast, some LMXBs have smooth orbital variations (e.g. 4U 1820-30 see \citealt{Stella+etal+1987}), that are also likely to be caused by the absorption of the accretion disk structure, but they are not classified as dipping sources.

To probe the orbital evolution of dipping LMXBs, the dip center time (usually determined as the time at which the minimum dip intensity occurs) is usually selected as the fiducial point. However, because the dip is caused by the absorption of the accretion disk structure, it is not completely phase locked with the orbital phase, but sometimes shows phase jitter (e.g. $\pm 0.05$ cycle for X 1916-053, see \citealt{Chou+etal+2001}). Both the dip center time and the dip profile changes with time (\citealt{Chou+etal+2001,Hu+etal+2008}). Furthermore, the dip even occasionally disappears at the expected time (\citealt{Smale+Wachter+1999,Chou+etal+2001}). Because of the variable dip profile, the fiducial point (the dip center time) is sometimes difficult to uniquely define. For example, \cite{Chou+etal+2001} obtained a dip center time by fitting a quadratic curve around the intensity minimum, whereas \cite{Homer+etal+2001} used Gaussian to model the dip profile, but neither approach is suitable for a dip with a complicated profile. \cite{Hu+etal+2008} introduced a method to extract the parameters of a dip, including the dip center time, dip width, and strength, regardless how complicated the dip profile is. With these parameters, the X-ray dip provides not only information about the orbital period evolution, but also the accretion disk dynamics as it is caused by the absorption of the structure of the accretion disk (\citealt{Chou+etal+2009}).

\begin{figure}
   \centering
   \includegraphics[width=14.0cm, angle=0]{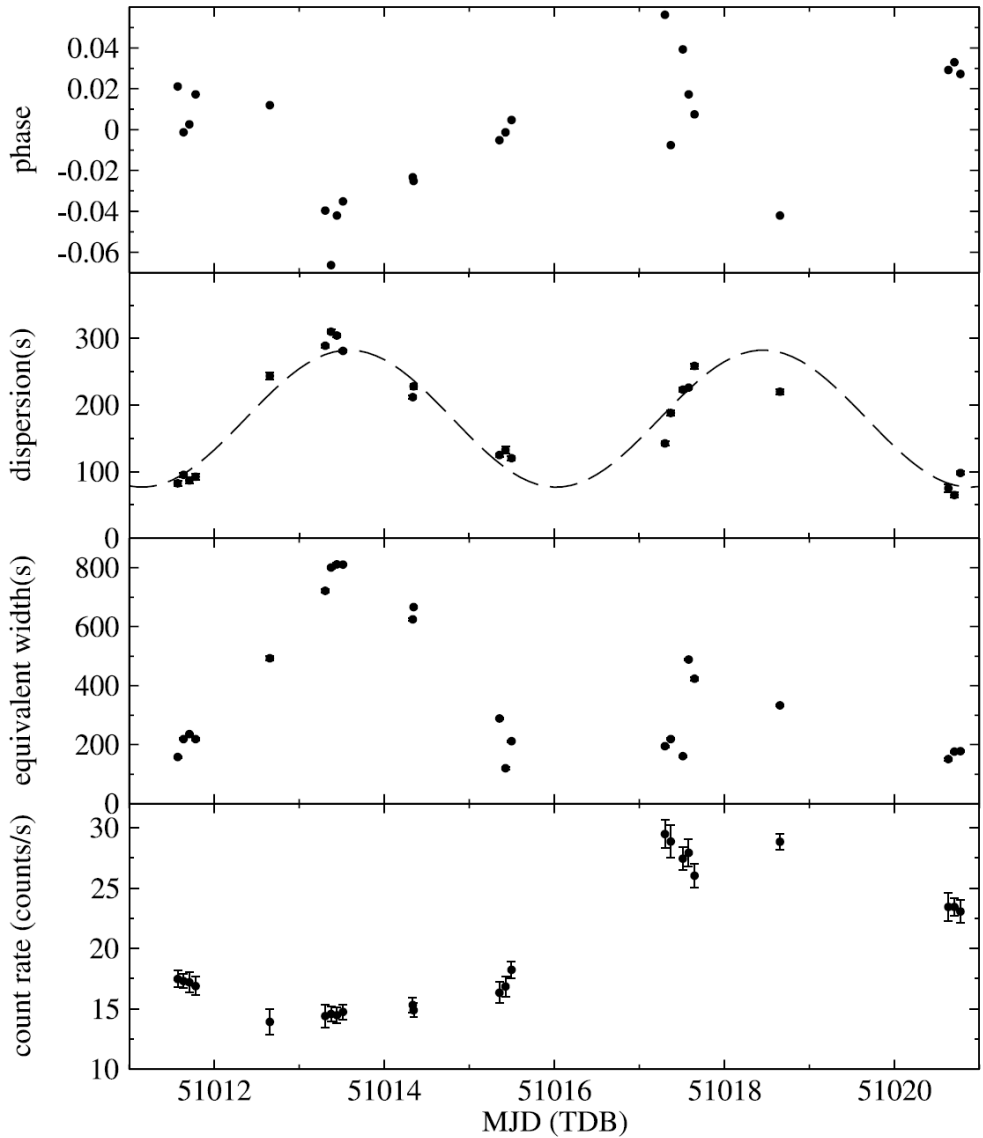}
   \caption{The dip parameters of X1916-053 from RXTE 1998 observations folded with a 4.87 d period. The dispersion (i.e. dip width) has a clear 4.87d periodicity. Adopted from  \cite{Hu+etal+2008}} 
   \label{hu2008}
   \end{figure}

X1916-053 is a typical dipping LMXB and exhibits many interesting timing phenomena.  It shows recurrent X-ray dips, first discovered by \cite{White+Swank+1982} and \cite{Walter+etal+1982} with a period of 3000 s. The secondary dip is sometimes seen with a phase difference of $\sim 0.5$ to the primary one. Other period in X-ray band ranging from 2985 s to 3015 s have been reported with marginal evidence (\citealt{Walter+etal+1982,White+Swank+1982,Smale+etal+1989,Yoshida+etal+1995,Church+etal+1997}). The most interesting point is that its optical counterpart modulates with a stable period of 3027 s, which is only $\sim 1$\% longer but significantly different to the 3000 s  X-ray dips (\citealt{Callanan+etal+1995}). Two different models to explain the discrepancy have been proposed. The $\sim 1$\% longer periodicity in the optical band compared to the X-ray band is likely caused by the coupling of a $\sim 3.9$d disk precession and the 3000 s orbital motion as the positive superhump seen in the SU UMa type dwarf nova. On the other hand, to explain the period of $\sim 199$d long-term modulation found by \cite{Priedhorsky+Terrell+1989} and the stable optical period (\citealt{Callanan+etal+1995,Chou+etal+2001}), \cite{Grindlay+1989,Grindlay+1992} proposed that X1916-053 is a hierarchical triple system. The long-term variation is due to eccentricity of the inner binary changes, which affects the accretion rate similar to the  $\sim 171$ d modulation seen in 4U 1820-30 (\citealt{Chou+Grindlay+2001}). The triple model predicts that the optical period (3027 s) is the orbital period whereas the SU UMa model suggests the X-ray dip period (3000 s) is the orbital period as in other dipping LMXBs. However, the $\sim 199$d long-term periodicity has never been confirmed by the subsequent observations (e.g. \citealt{Wen+etal+2006}). Neither have other predictions from the triple model been confirmed, except for the marginal evidence that the X-ray bursts occur clustered  with dips (\citealt{Chou+etal+2001}). On the other hand, although the optical period of X1916-053 is rather stable ($\dot P_{opt} \approx 10^{-10}$ s s$^{-1}$) in comparison to the superhumps of SU Ma systems ($\dot P_{sup} \sim 10^{-5}$ s s$^{-1}$), the X-ray dip period seems even more stable than the optical one (\citealt{Chou+etal+2001}). \cite{Chou+etal+2001} noted that the superhump may be stable in some CV systems (e.g. AM CVn, $\dot P_{sup} = 1.7 \times 10^{-11}$ s s$^{-1}$, \citealt{Solheim+etal+1998}). Therefore, the evidence is more favorable for the SU UMa model than the triple model. Furthermore, \cite{Retter+etal+2002} found the negative superhump with period of 2979s as a consequence of retrograde nodal precession of the accretion disk in a period of 4.8d coupling with orbital motion. \cite{Hu+etal+2008} discovered a 4.87d periodic variation in the dip width (Figure~\ref{hu2008}). These observations suggest that X1916-053 is the first LMXB to be classified as a permanent superhump system with both positive and negative superhump variations, like some CVs, such as AM CVn.

The configuration of CVs and LMXBs are very similar, even though the accretors and emission spectra are different.  The superhumps (at least the positive one) are believed to be caused by the presence of 3:1 resonance in the accretion disk (\citealt{Whitehurst+1988}). This resonance can only occur for the accretion with a mass ratio $q=m_{d}/m_{a} \lesssim 1/3$. This extreme value is more easily achieved in LMXBs than in CVs because of the larger accretors mass. Consequently, superhumps are expected in many LMXBs. In principle, superhumps may be seen in all the black hole LMXBs, as all verified stellar size black holes have mass larger than $3.2$ $M_{\sun}$  and the  masses of the donors are in general less massive than  1 $M_{\sun}$. Superhumps have been detected in a few black hole LMXBs during their X-ray outburst states, including GRO J0422+32, Nova Mus 1991, GS 2000+25 (\citealt{O'Donoghue+Charles+1996}) and XTE J1118+480 (\citealt{Zurita+etal+2002}). For LMXBs with neutron stars as accretors, \cite{Haswell+etal+2001} noted that superhumps are expected where the orbital period is less than 4.2h. In addition to X1916-053, a 693.5s positive superhump period was detected in the far-ultraviolet band by HST in 4U 1820-30, an ultra-compact LMXB with an orbital period of 685s (\citealt{Wang+Chakrabarty+2010}). Recently, \cite{Cornelisse+etal+2013} reported the discovery of a negative superhump in XB 1254-690, a dipping LMXB with orbital period of 3.9 h, in the optical band, as well as marginal evidence for the presence of a positive superhump. It appears reasonable that the most of the superhumps in LMXBs are detected in the longer wavelength bands as they reflect the behavior of the accretion disk. X1916-053 is currently the only case where the superhumps may be detected in the X-ray band. However, because the X-ray dips are caused by the absorption of the disk structure, they allow us to probe the disk precessions. Superhumps or related accretion disk precession phenomena are expected to be detected in other dipping LMXBs.

\section{Measuring orbital period by the X-ray pulsar}
\label{sect:OXP}

If one of the components of a binary system is a pulsar, the orbital period and other orbital parameters can be obtained from the orbital Doppler effect of the pulsar, or more precisely from the pulse arrival time delay due to orbital motion. However, the pulsation was rarely seen in neutron star LMXBs before 1998, probably because the weak magnetic field ($\sim 10^8-10^9$ G) was ``buried'' or ``screened'' by the accreted matter so the accretion flow cannot be channeled onto the magnetic poles (\citealt{Cumming+etal+2001}), or due to the insufficient sensitivities of X-ray telescopes. The only cases were 4U 1626-67 ($P_s = 7.67$ s, \citealt{Rappaport+etal+1977}), Her X-1 ($P_s = 1.24$ s, \citealt{Tananbaum+etal+1972}), GX 1+4 ($P_s \approx 2.3$ min, \citealt{Lewin+etal+1971}) and GRO J1744-28 ($P_s =0.467$ s \citealt{Finger+etal+1996}). The accretor of the LMXB identified as a neutron star was usually determined from its exhibition of an X-ray burst, a thermonuclear explosion on the surface of the neutron star. The LMXBs were considered as the progenitors of the radio millisecond pulsars ($P_s \lesssim 10$ ms, \citealt{Alpar+etal+1982,Radhakrishnan+etal+1982}), but no millisecond pulsation had ever been detected in LMXBs by 1998, even though there were some indirect evidences that many of the neutron stars in LMXBs spin at a frequency of hundreds of Hz, like the quasi-periodic oscillation (OPQ.) during the X-ray burst (called burst QPO). The first accreting millisecond X-ray pulsar (AXMP) SAX 1808.4-3658 was discovered in 1998 by the {\it Rossi X-ray Timing Explorer (RXTE)}, an X-ray telescope with large X-ray photon collecting area and unprecedented time resolution (1 $\mu$s) (\citealt{Wijnands+vanderKlis+1998}). Its 2 h orbital period was immediately determined using the pulse arrival time delay technique (\citealt{Chakrabarty+Morgan+1998}). To date, a total of 15 AXMPs have been discovered. Of these, 14 of them have had their orbital periods measured precisely using their pulsations; the exception is Aql X-1 where the pulsation was detected for only 150s (\citealt{Casella+etal+2008}), which is too short to determine the orbital parameters for this 18.95 h orbital period binary. Furthermore, the frequency of burst QPO has been proven to be identical (or doubled) to the neutron star frequency (\citealt{Chakrabarty+etal+2003}), which provides us an alternative way to measure the neutron star spin period in an LMXB. As a result, the burst QPO sources with QPO frequencies of hundreds of Hz are also called ``nuclear-powered millisecond pulsars''  (\citealt{Chakrabarty+etal+2003}). 

However, not all LMXBs with pulsations can have orbital periods determined through the orbital Doppler effect or the pulse arrival delay technique. The ultra-compact LMXB 4U 1626-67, with a pulsation period of 7.67 s and orbital period of 0.7 h, detected by the sidebands of optical pulsation (\citealt{Middleditch+etal+1981}), shows neither a periodic Doppler shift nor a pulse arrival delay caused by the orbital motion due to its very small projected semi-major axis ($a_x$ sin {\it i}$<$ 0.04 lt-s, \citealt{Middleditch+etal+1981}).  In addition, the orbital parameters are almost impossible to measure for systems where the pulsation is only detectable for a short time, such as Aql X-1 (150 s vs. 19 h orbital period) and burst QPOs (several seconds vs. orbital period of tens of minutes to several hours). Otherwise, the pulsation provides a way to very precisely measure the orbital parameters, as well as the neutron star spin parameters. Higher pulsation frequency can give higher precision orbital and spin parameters, which allow us to measure the evolution of these parameters, such as orbital and spin period changes, in short observational time span.  For example, we consider an analysis technique similar to the O-C method to measure the spin period of a millisecond pulsar with a spin frequency about 100 Hz. If the deviation between the true spin period (observed) and the guess period (calculated) is only a million to one, the pulse phase will have significant drift in less than $10^6$ cycles, which corresponds to less than 10 ks. In other words, the spin period can be measured with an error of less than $10^{-6}$ in a time span 3 hours.

\begin{figure}
   \centering
   \includegraphics[width=14.0cm, angle=0]{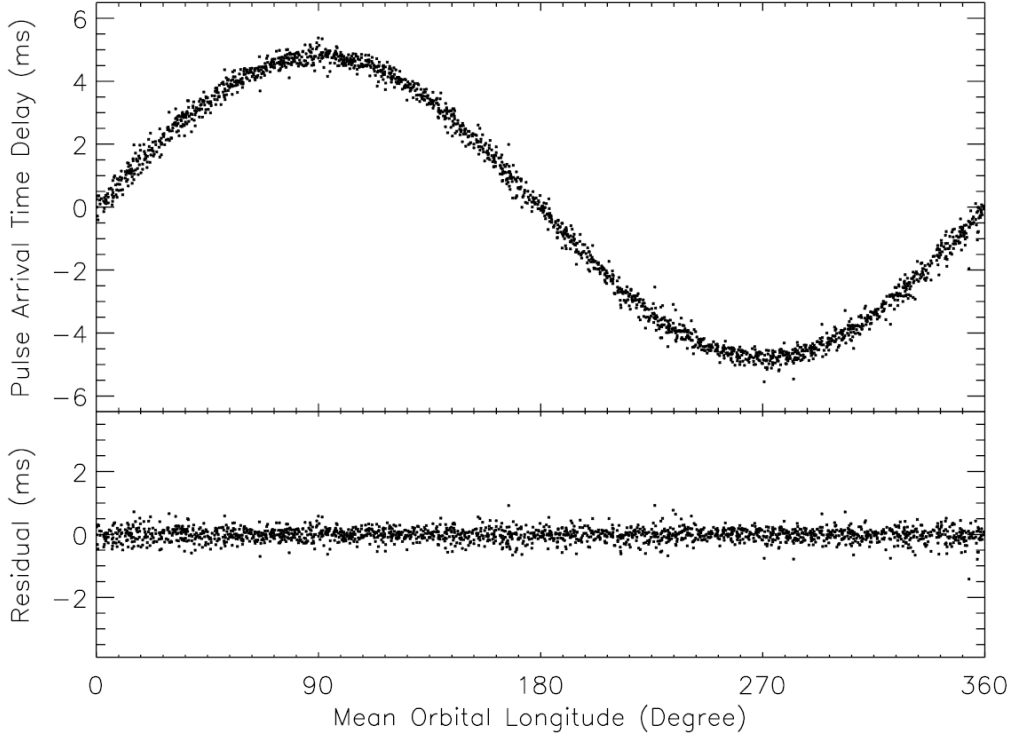}
   \caption{Pulse arrival time delay due to orbital motion observed from XTE J1807-294. The orbital period and other orbital parameters (e.g., the projected radius) can be detected from this technique. Adopted from \cite{Chou+etal+2008}} 
   \label{chou2008-0}
   \end{figure}

For a pulsar in a binary system, the phase drift (or pulse arrival time delay) may be caused by orbital motion (see Figure~\ref{chou2008-0}) as well as the neutron star spin frequency change (if the observation time span is long enough). Because the LMXB is an old system, the orbital eccentricity is almost always close to zero. On the other hand, the spin frequency evolution is usually modeled as a polynomial. If the effect of the frequency derivative is small, the observed frequency of pulsation can be written as

\begin{eqnarray}\label{orb_dop}
\nu(t) &=& \nu_0 {{2\pi(a_x \sin i)} \over {cP_{orb}}} \sin \biggl[ {{2\pi(t-T_{\pi/2})} \over P_{orb}}  \biggr]+\nu_0+\dot \nu (t-T_0)+{1 \over 2}\dot\dot \nu(t-T_0)^2+......\\
\nonumber & = & A \nu_0 f_{orb} \sin \bigl [ {{2\pi f_{orb} (t-T_{\pi/2})} } \bigr] 
+\nu_0+\dot \nu (t-T_0)+{1 \over 2}\ddot \nu(t-T_0)^2+......
\end{eqnarray}  

\noindent where $P_{orb}$ is the orbital period, $f_{orb}$ is the orbital frequency, $a_x$ is the orbital radius of the neutron star, $i$ is the inclination angle, $A \equiv a_x \sin i / c$, $T_{\pi/2}$ is a reference time when the neutron star is at superior conjunction and $T_0$ is a reference time for the pulsation, which usually takes the time of a specified fiducial point of pulsation. The cycle count can be evaluated as in the definition in Equation~\ref{oc1}. The phase is defined as the fractional part of the cycle count. Because the X-ray detector records the event arrival time, the phase can be defined on each X-ray event as
 \begin{eqnarray}\label{orb_cyc}
\phi_i &=& frac\biggl(\int_{T_0}^{t_i} \nu(t^\prime)dt^{\prime} \biggr)\\
\nonumber & = & frac \biggl\{-A \nu_0  \cos \bigl [ {{2\pi f_{orb} (t-T_{\pi/2})} } \bigr] + A \nu_0  \cos \bigl [ {{2\pi f_{orb} (T_0-T_{\pi/2})} } \bigr] \\
\nonumber & & +\nu_0(t_i-T_0)+{1 \over 2}\dot \nu(t_i-T_0)^2+{1 \over 6}\ddot \nu(t_i-T_0)^3+...... \biggr\}
\end{eqnarray}

\noindent where $t_i$ is the event arrival time of the ith X-ray photon. The pulse profile can be obtained by binning the event phases. 

However, if the parameters applied to Equation~\ref{orb_cyc} deviate from the true values, significant phase drift can be introduced in the evolution of the pulse profile.  Therefore, we may adopt the method analogous to the O-C method to refine the parameters. If the deviations between guess and true parameters are small, the phase drift of the pulse profile can be expressed (to first order approximation) as

\begin{eqnarray}\label{orb_corr}
\delta \phi(t) &=&  \bigl \{  -(t-T_0^{(0)})+A^{(0)} \cos \bigl [ {{2\pi f_{orb}^{(0)} (t-T_{\pi/2}^{(0)})} } \bigr]  - A^{(0)} \cos \bigl [ {{2\pi f_{orb}^{(0)} (T_0^{(0)}-T_{\pi/2}^{(0)})} } \bigr]\bigr \} \delta \nu_0 \\
\nonumber & & + \bigl \{ \nu_0^{(0)}+2\pi A^{(0)} \nu_0^{(0)}f_{orb}^{(0)}\cos \bigl [ {{2\pi f_{orb}^{(0)} (T_0^{(0)}-T_{\pi/2}^{(0)})} } \bigr]     \bigr\} \delta T_0\\
\nonumber & & +\bigl \{ \nu_0^{(0)}\cos \bigl [ {{2\pi f_{orb}^{(0)} (t-T_{\pi/2}^{(0)})} } \bigr] - \nu_0^{(0)} \cos \bigl [ {{2\pi f_{orb}^{(0)} (T_0^{(0)}-T_{\pi/2}^{(0)})} } \bigr] \bigr \} \delta A \\
\nonumber & & + \bigl \{-2\pi\nu_0^{(0)}A^{(0)}(t-T_{\pi/2}^{(0)}) \sin  \bigl [ {{2\pi f_{orb}^{(0)} (t-T_{\pi/2}^{(0)})} } \bigr] + 2\pi\nu_0^{(0)}A^{(0)}(T_0^{(0)}-T_{\pi/2}^{(0)}) \sin  \bigl [ {{2\pi f_{orb}^{(0)} (T_0^{(0)}-T_{\pi/2}^{(0)})} } \bigr]\bigr \} \delta f_{orb}\\
\nonumber & & + \bigl \{ 2\pi \nu_0^{(0)}A^{(0)} f_{orb}^{(0)}  \sin  \bigl [ {{2\pi f_{orb}^{(0)} (t-T_{\pi/2}^{(0)})} } \bigr] - 2\pi \nu_0^{(0)}A^{(0)} f_{orb}^{(0)}  \sin  \bigl [ {{2\pi f_{orb}^{(0)} (T_0^{(0)}-T_{\pi/2}^{(0)})} } \bigr]\bigr \} \delta T_{\pi/2}^{(0)}\\
\nonumber & & +\bigl [ - {1 \over 2}(t-T_0^{(0)})^2 \bigr ] \delta \dot \nu +\bigl [ - {1 \over 6}(t-T_0^{(0)})^3 \bigr ] \delta {\ddot \nu} +......
\end{eqnarray}

\noindent where the parameters with superscripts (0) indicate the guess parameters and the parameters with $\delta$ are the differences between true and guess parameters. The parameter corrections can be obtained by fitting the phase drift with the time. Although Equation~\ref{orb_corr} is complex, it is basically a linear fitting with exact solution. The initial orbital parameters and pulsar frequency may be obtained by the orbital Doppler shift from the power spectra, and the initial pulsar frequency derivative may be set to zero if the data time span is not very large. This correction process can be repeated until the corrections of parameters are much smaller than the corresponding errors from the fitting. 

\begin{figure}
   \centering
   \includegraphics[width=14.0cm, angle=0]{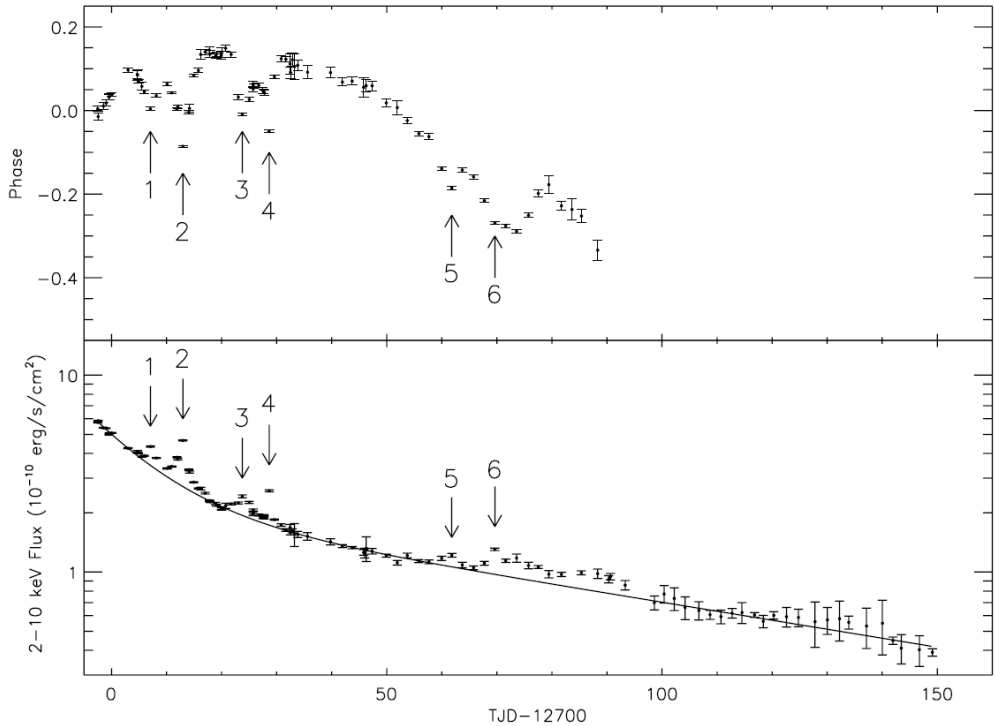}
   \caption{Pulse phase vs. X-ray flux of XTE J1807-294 from its 2003 outburst. The negative phase drops are seen to be coincident with the X-ray flux enhancements. Adopted from \cite{Chou+etal+2008}} 
   \label{chou2008-1}
   \end{figure}

In principle, there is no pulse phase drift either from orbital motion or from the secular term (i.e. $[-1/2(t-T_0^{(0)})^2 ]\delta \dot \nu +[-1/6(t-T_0^{(0)})^3]\delta {\ddot \nu} ......$ in Equation~\ref{orb_corr}) when the best orbital and spin parameters are applied to the folding. In other words, all the fiducial points of pulsations should be aligned at phase zero. However, this simple model is not usually applicable to AXMPs because the secular term is too complicated to model as a polynomial.  Such abnormal secular phase variation, also called timing noise, is believed to be due to accretion rate changes (\citealt{Patruno+etal+2009}). A typical example is the pulse variation of the AXMP XTE J1807-294 during its 2003 outburst. \cite{Chou+etal+2008} discovered that the pulse phase exhibited a large phase drop 6 times during the $\sim$90 d when the pulsation was detectable. These large phase drops are too large to be explained by any known accretion torque theories. However, by comparing with the X-ray light curve, these phase drops are coincident with the X-ray flares (Figure~\ref{chou2008-1}). Furthermore, the fractional pulse amplitude is also enhanced during the flare. \cite{Chou+etal+2008} noted that these observations imply that the accreting hot spot shifts as the accretion rate changes during the X-ray flare. Therefore, although the AXMP is a good candidate for studying the accretion torque because of its small moment of inertia ($I_{NS} \approx 10^{-5}I_{WD}$) and high spin frequency (allowing precise measurement of parameters), the timing noise problem due to accretion rate variation must be solved in advanced. Nevertheless, precise orbital parameters are still essential for this study. The problem is that there is no good model to describe the timing noise. Fortunately, the time scale of the timing noise is usually much longer than the orbital period in AXMPs. This phase drift can be removed locally by modeling it as a linear trend for a short time span of data and obtaining the orbital parameters (see \citealt{Chou+etal+2008} for more detail).  

\section{Summary}
\label{sect:SUM}

X-ray binaries have been one of the most important astrophysical sources in the X-ray band since the beginning of X-ray astronomy in the early 1960'.  The LMXBs, with binary separations on the order of a solar radius, allow us to study the nature of compact objects, accretion phenomena, accretion disk dynamics and the evolution of accreting or close binary systems. The orbital period is the fundamental parameter for a binary system. In addition to the optical band, X-ray emissions provide another way to measure the orbital period of LMXB systems. To date, more than 30 years of X-ray observation data have been archived. Because it is generally believed that the orbital period derivative for a LMXB system is $\dot P_{orb}/P_{orb} \sim 10^{-7} - 10^{-8}$ yr$^{-1}$, the orbital period derivatives for some of the LMXBs with orbital periods of hours may be resolved from the historical data by the O-C method discussed in Section~\ref{sect:OCM}. This allows us to learn more about the accreting binary evolution and accretion mechanism. Furthermore, some LMXBs with very short orbital periods ($< 80$ min), called ultra-compact X-ray binaries (UCXBs), e.g. 4U 1820-30 ($P_{orb}=11.5$ min), XTE J1807-294 ($P_{orb}=40.5$ min) and X 1916-053 ($P_{orb}=50$ min), whose mass-losing companions are  hydrogen-deficient degenerate stars (\citealt{Rappaport+etal+1982}), are good candidates for probing the gravitational wave radiations, in addition to the AM CVn systems in the future.    
\begin{acknowledgements}

I am grateful for helpful comments on this paper by Craig Heinke and the anonymous referee. This work is partially supported by Taiwan Ministry of Science and Technology grant NSC 102-2112-M-008-020-MY3.

\end{acknowledgements}

\bibliographystyle{raa}
\bibliography{ms1812bibtex}

\begin{thebibliography}{64}
\providecommand{\natexlab}[1]{#1}
\providecommand{\selectlanguage}[1]{\relax}

\bibitem[{{Alpar} et~al.(1982){Alpar}, {Cheng}, {Ruderman}, \&
  {Shaham}}]{Alpar+etal+1982}
{Alpar}, M.~A., {Cheng}, A.~F., {Ruderman}, M.~A., \& {Shaham}, J. 1982, \nat,
  300, 728

\bibitem[{{Buccheri} et~al.(1983){Buccheri}, {Bennett}, {Bignami}
  et~al.}]{Buccheri+etal+1983}
{Buccheri}, R., {Bennett}, K., {Bignami}, G.~F., et~al. 1983, \aap, 128, 245

\bibitem[{{Callanan} et~al.(1995){Callanan}, {Grindlay}, \&
  {Cool}}]{Callanan+etal+1995}
{Callanan}, P.~J., {Grindlay}, J.~E., \& {Cool}, A.~M. 1995, \pasj, 47, 153

\bibitem[{{Casella} et~al.(2008){Casella}, {Altamirano}, {Patruno}, {Wijnands},
  \& {van der Klis}}]{Casella+etal+2008}
{Casella}, P., {Altamirano}, D., {Patruno}, A., {Wijnands}, R., \& {van der
  Klis}, M. 2008, \apjl, 674, L41

\bibitem[{{Chakrabarty} \& {Morgan}(1998)}]{Chakrabarty+Morgan+1998}
{Chakrabarty}, D., \& {Morgan}, E.~H. 1998, \nat, 394, 346

\bibitem[{{Chakrabarty} et~al.(2003){Chakrabarty}, {Morgan}, {Muno}
  et~al.}]{Chakrabarty+etal+2003}
{Chakrabarty}, D., {Morgan}, E.~H., {Muno}, M.~P., et~al. 2003, \nat, 424, 42

\bibitem[{{Chou} et~al.(2008){Chou}, {Chung}, {Hu}, \& {Yang}}]{Chou+etal+2008}
{Chou}, Y., {Chung}, Y., {Hu}, C.-P., \& {Yang}, T.-C. 2008, \apj, 678, 1316

\bibitem[{{Chou} \& {Grindlay}(2001)}]{Chou+Grindlay+2001}
{Chou}, Y., \& {Grindlay}, J.~E. 2001, \apj, 563, 934

\bibitem[{{Chou} et~al.(2001){Chou}, {Grindlay}, \& {Bloser}}]{Chou+etal+2001}
{Chou}, Y., {Grindlay}, J.~E., \& {Bloser}, P.~F. 2001, \apj, 549, 1135

\bibitem[{{Chou} et~al.(2009){Chou}, {Hu}, {Chung}, \& {Yang}}]{Chou+etal+2009}
{Chou}, Y., {Hu}, C.-P., {Chung}, Y., \& {Yang}, T.-C. 2009, in Conference
  Proceedings of 10th Asian-Pacific Regional IAU Meeting, edited by S.-N.
  {Zhang}, L.~{Yan}, \& Q.~{Yu}, 210--211 (Kunming, China)

\bibitem[{{Church} et~al.(1997){Church}, {Dotani}, {Balucinska-Church}
  et~al.}]{Church+etal+1997}
{Church}, M.~J., {Dotani}, T., {Balucinska-Church}, M., et~al. 1997, \apj, 491,
  388

\bibitem[{{Cornelisse} et~al.(2013){Cornelisse}, {Kotze}, {Casares}, {Charles},
  \& Hakala}]{Cornelisse+etal+2013}
{Cornelisse}, R., {Kotze}, M.~M., {Casares}, J., {Charles}, P.~A., \& Hakala,
  P.~J. 2013, \mnras, 436, 910

\bibitem[{{Cumming} et~al.(2001){Cumming}, {Zweibel}, \&
  {Bildsten}}]{Cumming+etal+2001}
{Cumming}, A., {Zweibel}, E., \& {Bildsten}, L. 2001, \apj, 557, 958

\bibitem[{{de Jager} et~al.(1989){de Jager}, {Raubenheimer}, \&
  {Swanepoel}}]{deJager+etal+1989}
{de Jager}, O.~C., {Raubenheimer}, B.~C., \& {Swanepoel}, J. W.~H. 1989, \aap,
  221, 180

\bibitem[{{Finger} et~al.(1996){Finger}, {Koh}, {Nelson}
  et~al.}]{Finger+etal+1996}
{Finger}, M.~H., {Koh}, D.~T., {Nelson}, R.~W., et~al. 1996, \nat, 381, 291

\bibitem[{{Frank} et~al.(1987){Frank}, {King}, \& {Lasota}}]{Frank+etal+1987}
{Frank}, J., {King}, A.~R., \& {Lasota}, J.~P. 1987, \aap, 178, 137

\bibitem[{{Grindlay}(1989)}]{Grindlay+1989}
{Grindlay}, J.~E. 1989, in in Proc. 23rd ESLAB Symp. on Two Topics in X-Ray
  Astronomy, \emph{(ESA SP-296; Noordwijk: ESA)}, vol.~1, edited by J.~{Hunt}
  \& B.~{Battrick}, 121--126

\bibitem[{{Grindlay}(1992)}]{Grindlay+1992}
{Grindlay}, J.~E. 1992, in in Proc. 28th Yamada Conf., Frontiers of X-Ray
  Astronomy, \emph{(Tokyo: Universal Acadamy),}, vol.~1, edited by Y.~{Tanaka}
  \& K.~{Koyama}, 69

\bibitem[{{Haswell} et~al.(2001){Haswell}, {King}, {Murray}, \&
  {Charles}}]{Haswell+etal+2001}
{Haswell}, C.~A., {King}, A.~R., {Murray}, J.~R., \& {Charles}, P.~A. 2001,
  \mnras, 321, 475

\bibitem[{{Heinke} et~al.(2005){Heinke}, {Grindlay}, \&
  {Edmonds}}]{Heinke+etal+2005}
{Heinke}, C.~O., {Grindlay}, J.~E., \& {Edmonds}, P.~D. 2005, \apj, 622, 556

\bibitem[{{Heinke} et~al.(2003){Heinke}, {Grindlay}, {Lloyd}, \&
  {Edmonds}}]{Heinke+etal+2003}
{Heinke}, C.~O., {Grindlay}, J.~E., {Lloyd}, D.~A., \& {Edmonds}, P.~D. 2003,
  \apj, 588, 452

\bibitem[{{Hertz} et~al.(1997){Hertz}, {Wood}, \& {Cominsky}}]{Hertz+etal+1997}
{Hertz}, P., {Wood}, K.~S., \& {Cominsky}, L.~R. 1997, \apj, 486, 1000

\bibitem[{{Homer} et~al.(2001){Homer}, {Charles}, {Hakala}
  et~al.}]{Homer+etal+2001}
{Homer}, L., {Charles}, P.~A., {Hakala}, P., et~al. 2001, \mnras, 322, 827

\bibitem[{{Horne} \& {Baliunas}(1986)}]{Horne+Baliunas+1986}
{Horne}, J.~H., \& {Baliunas}, S.~L. 1986, \apj, 302, 757

\bibitem[{{Hu} et~al.(2008){Hu}, {Chou}, \& {Chung}}]{Hu+etal+2008}
{Hu}, C.-P., {Chou}, Y., \& {Chung}, Y. 2008, \apj, 680, 1405

\bibitem[{{Iaria} et~al.(2011){Iaria}, {di Salvo}, {Burderi}
  et~al.}]{Iaria+etal+2011}
{Iaria}, R., {di Salvo}, T., {Burderi}, L., et~al. 2011, \aap, 534

\bibitem[{{Jain} \& {Paul}(2011)}]{Jain+Paul+2011}
{Jain}, C., \& {Paul}, B. 2011, \mnras, 183, 413

\bibitem[{{Jonker} \& {van der Klis}(2001)}]{Jonker+vanderKlis+2001}
{Jonker}, P.~G., \& {van der Klis}, M. 2001, \apjl, 553, L43

\bibitem[{{Levine} et~al.(2011){Levine}, {Bradt}, {Chakrabarty}, {Corbet}, \&
  {Harris}}]{Levine+etal+2011}
{Levine}, A.~M., {Bradt}, H.~V., {Chakrabarty}, D., {Corbet}, R. H.~D., \&
  {Harris}, R.~J. 2011, \apjs, 196, 6

\bibitem[{{Lewin} et~al.(1971){Lewin}, {Ricker}, \&
  {McClintock}}]{Lewin+etal+1971}
{Lewin}, W. H.~G., {Ricker}, G.~R., \& {McClintock}, J.~E. 1971, \apjl, 169,
  L17

\bibitem[{{Liu} et~al.(2006){Liu}, {van Paradijs}, \& {van den
  Heuvel}}]{Liu+etal+2006}
{Liu}, Q.~Z., {van Paradijs}, J., \& {van den Heuvel}, E. P.~J. 2006, \aap,
  455, 1165

\bibitem[{{Liu} et~al.(2007){Liu}, {van Paradijs}, \& {van den
  Heuvel}}]{Liu+etal+2007}
{Liu}, Q.~Z., {van Paradijs}, J., \& {van den Heuvel}, E. P.~J. 2007, \aap,
  469, 807

\bibitem[{{Lomb}(1976)}]{Lomb+1976}
{Lomb}, N.~R. 1976, \apss, 39, 447

\bibitem[{{Maeda} et~al.(2013){Maeda}, {Mori}, \& {Dotani}}]{Maeda+etal+2013}
{Maeda}, Y., {Mori}, H., \& {Dotani}, T. 2013, Adv Space Res, 51, 1278

\bibitem[{{Middleditch} et~al.(1981){Middleditch}, {Mason}, {Nelson}, \&
  {White}}]{Middleditch+etal+1981}
{Middleditch}, J., {Mason}, K.~O., {Nelson}, J.~E., \& {White}, N.~E. 1981,
  \apj, 244, 1001

\bibitem[{{Nowak} et~al.(2002){Nowak}, {Heinz}, \&
  {Begelman}}]{Nowak+etal+2002}
{Nowak}, M.~A., {Heinz}, S., \& {Begelman}, M.~C. 2002, \apj, 573, 778

\bibitem[{{O'Donoghue} \& {Charles}(1996)}]{O'Donoghue+Charles+1996}
{O'Donoghue}, D., \& {Charles}, P.~A. 1996, \mnras, 282, 191

\bibitem[{{Parmar} et~al.(2000){Parmar}, {Oosterbroek}, {Del Sordo}
  et~al.}]{Parmar+etal+2000}
{Parmar}, A.~N., {Oosterbroek}, T., {Del Sordo}, S., et~al. 2000, \aap, 356,
  175

\bibitem[{{Parmar} et~al.(1991){Parmar}, {Smale}, {Verbunt}, \&
  {Corbet}}]{Parmar+etal+1991}
{Parmar}, A.~N., {Smale}, A.~P., {Verbunt}, F., \& {Corbet}, R. H.~D. 1991,
  \apj, 366, 253

\bibitem[{{Patruno} et~al.(2009){Patruno}, {Wijnands}, \& {van der
  Klis}}]{Patruno+etal+2009}
{Patruno}, A., {Wijnands}, R., \& {van der Klis}, M. 2009, \apjl, 698, L60

\bibitem[{{Priedhorsky} \& {Terrell}(1984)}]{Priedhorsky+Terrell+1989}
{Priedhorsky}, W.~C., \& {Terrell}, J. 1984, \apj, 280, 661

\bibitem[{Radhakrishnan \& Srinivasan(1982)}]{Radhakrishnan+etal+1982}
Radhakrishnan, V., \& Srinivasan, G. 1982, \sci, 51, 1096

\bibitem[{{Rappaport} et~al.(1982){Rappaport}, {Joss}, \&
  {Webbink}}]{Rappaport+etal+1982}
{Rappaport}, S., {Joss}, P.~C., \& {Webbink}, R.~F. 1982, \apj, 254, 616

\bibitem[{{Rappaport} et~al.(1977){Rappaport}, {Markert}, {Li}
  et~al.}]{Rappaport+etal+1977}
{Rappaport}, S., {Markert}, T., {Li}, F.~K., et~al. 1977, \apjl, 217, L29

\bibitem[{{Retter} et~al.(2002){Retter}, {Chou}, {Bedding}, \&
  {Naylor}}]{Retter+etal+2002}
{Retter}, A., {Chou}, Y., {Bedding}, T.~R., \& {Naylor}, T. 2002, \mnras, 330,
  L37

\bibitem[{{Scargle}(1982)}]{Scargle+1982}
{Scargle}, J.~D. 1982, \apj, 263, 835

\bibitem[{{Smale} et~al.(1989){Smale}, {Mason}, {Williams}, \&
  {Watson}}]{Smale+etal+1989}
{Smale}, A.~P., {Mason}, K.~O., {Williams}, O.~R., \& {Watson}, M.~G. 1989,
  \pasj, 41, 607

\bibitem[{{Smale} \& {Wachter}(1999)}]{Smale+Wachter+1999}
{Smale}, A.~P., \& {Wachter}, S. 1999, \apj, 527, 341

\bibitem[{{Solheim} et~al.(1998){Solheim}, {Provencal}, {Bradley}
  et~al.}]{Solheim+etal+1998}
{Solheim}, J.-E., {Provencal}, J.~L., {Bradley}, P.~A., et~al. 1998, \aap, 332,
  939

\bibitem[{{Stella} et~al.(1987){Stella}, {Priedhorsky}, \&
  {White}}]{Stella+etal+1987}
{Stella}, L., {Priedhorsky}, W., \& {White}, N.~E. 1987, \apjl, 312, L17

\bibitem[{{Stellingwerf}(1978)}]{Stellingwerf+1978}
{Stellingwerf}, R.~F. 1978, \apj, 224, 953

\bibitem[{{Tananbaum} et~al.(1972){Tananbaum}, {Gursky}, {Kellogg}
  et~al.}]{Tananbaum+etal+1972}
{Tananbaum}, H., {Gursky}, H., {Kellogg}, E.~M., et~al. 1972, \apjl, 174, L143

\bibitem[{{Walter} et~al.(1982){Walter}, {Mason}, {Clarke}
  et~al.}]{Walter+etal+1982}
{Walter}, F.~M., {Mason}, K.~O., {Clarke}, J.~T., et~al. 1982, \apjl, 253, L67

\bibitem[{{Wang} \& {Chakrabarty}(2010)}]{Wang+Chakrabarty+2010}
{Wang}, Z., \& {Chakrabarty}, D. 2010, \apj, 712, 653

\bibitem[{{Wen} et~al.(2006){Wen}, {Levine}, {Corbet}, \&
  {Bradt}}]{Wen+etal+2006}
{Wen}, L., {Levine}, A.~M., {Corbet}, R. H.~D., \& {Bradt}, H.~V. 2006, \apjs,
  163, 372

\bibitem[{{White} et~al.(1981){White}, {Becker}, {Boldt}
  et~al.}]{White+etal+1981}
{White}, N.~E., {Becker}, R.~H., {Boldt}, E.~A., et~al. 1981, \apj, 247, 994

\bibitem[{{White} \& {Holt}(1982)}]{White+Holt+1982}
{White}, N.~E., \& {Holt}, S.~S. 1982, \apj, 257, 318

\bibitem[{{White} \& {Swank}(1982)}]{White+Swank+1982}
{White}, N.~E., \& {Swank}, J.~H. 1982, \apjl, 253, L61

\bibitem[{{Whitehurst}(1988)}]{Whitehurst+1988}
{Whitehurst}, R. 1988, \mnras, 232, 35

\bibitem[{{Wijnands} \& {van der Klis}(1998)}]{Wijnands+vanderKlis+1998}
{Wijnands}, R., \& {van der Klis}, M. 1998, \nat, 394, 344

\bibitem[{{Wolff} et~al.(2009){Wolff}, {Ray}, {Wood}, \&
  {Hertz}}]{Wolff+etal+2009}
{Wolff}, M.~T., {Ray}, P.~S., {Wood}, K.~S., \& {Hertz}, P.~L. 2009, \apjs,
  183, 156

\bibitem[{{Wolff} et~al.(2007){Wolff}, {Wood}, \& {Ray}}]{Wolff+etal+2007}
{Wolff}, M.~T., {Wood}, K.~S., \& {Ray}, P.~S. 2007, \apj, 668, L151

\bibitem[{{Yoshida} et~al.(1995){Yoshida}, {Inoue}, {Mitsuda}, {Dotani}, \&
  {Makino}}]{Yoshida+etal+1995}
{Yoshida}, K., {Inoue}, H., {Mitsuda}, K., {Dotani}, T., \& {Makino}, F. 1995,
  \pasj, 47, 141

\bibitem[{{Zurita} et~al.(2002){Zurita}, {Casares}, {Shahbaz}
  et~al.}]{Zurita+etal+2002}
{Zurita}, C., {Casares}, J., {Shahbaz}, T., et~al. 2002, \mnras, 333, 791

\end{thebibliography}

\end{document}